\documentstyle[twoside,fleqn,espcrc2]{article}

\newcommand{\AmS}{{\protect\the\textfont2
  A\kern-.1667em\lower.5ex\hbox{M}\kern-.125emS}}

\title{Some effects of the anisotropy in a simple lattice gauge model
       at finite temperature}

\author{L.~A.~Averchenkova, V.~K.~Petrov
        \address{Bogolyubov Institute for Theoretical Physics,\\
        National Academy of Sciences of Ukraine, Kiev 143, UKRAINE}}
       
\begin{document}

\begin{abstract}
Monte Carlo simulations are carried out on the $(3+1)$-dimensional Z(2)
anisotropic lattice model, and a new method to simulate extremely anisotropic
lattice systems with discrete symmetries is proposed. Dependence of the
temporal and spatial average plaquette, Wilson loops on the anisotropy
parameter is presented. 
\end{abstract}

\maketitle

\section{Introduction}

In our previous paper \cite{averch}, the phase structure for Z(N)
pure gauge theory with the lattice anisotropy $\xi\equiv a_\sigma/a_\tau$
\footnote{
The bare anisotropy
$\widetilde{\xi}\equiv\sqrt{\kappa_\tau/\kappa_\sigma}$,
$\kappa_\sigma(\kappa_\tau)$ denoting the spatial (temporal) gauge couplings,
is often considered as a proper lattice anisotropy parameter
instead of $\xi$.}
[$a_\sigma(a_\tau)$ is the spatial (temporal) spacing]
has been studied analytically. It has been particularly indicated there
that arbitrarily chosen trajectories in anisotropy parameter and gauge
coupling space can cross phase transition lines, and give rise to
discontinuities in thermodynamic quantities along those trajectories.

The Z(N) gauge theory in four dimensions was first analyzed numerically
for the Wilson action in a classical paper by Creutz, Jacobs and Rebbi
\cite{creutz}. Numerous Monte Carlo studies concern various aspects of
Z(N) gauge theories, i.e. order of phase transitions both in pure
gauge theories \cite{creutz} and in Z(N) gauge--Higgs models
\cite{jong,fern,gavai}, dependence of the phase picture on d \cite{jong}
and on N \cite{creutz,fern,bhanot}, elaboration of improved algorithms
to simulate such systems with metastable states \cite{bhanot,parisi}.
Our paper is aimed at clarifying the role of the lattice anisotropy
$\widetilde{\xi}$ in the phase structure description of the finite 
temperature 4d Z(2) gauge theory by using Monte Carlo technique.
Here, we check the analytic results of \cite{averch}
and picture the phase plane $[\kappa_\tau;\kappa_\sigma]$ in the area
we cannot investigate analytically.

\section{Algorithm}

The partition function is a sum over all configurations of the system
\begin{eqnarray}
{\cal Z} &=& \sum_{\{\sigma_{x;\nu}\}}\exp(-\kappa_{\nu\mu}{\cal S});
\label{1} \\
{\cal S} &=& -\sum_{x;\nu,\mu} {\rm Re}
(\sigma_{x;\nu}\sigma_{x+\nu;\mu}\sigma_{x+\mu;\nu}^*\sigma_{x;\mu}^*);
\nonumber \\
\sigma_{x;\nu} &=& \exp\left(2\pi iq_{x;\nu}/N\right) \in Z(N);
\nonumber \\
q_{x;\nu} &=& 0,\cdots,N-1.
\nonumber
\end{eqnarray}
Although this is a finite sum, the number of configurations
$\{\sigma_{x;\nu}\}$ is so large
even for lattices which are only a few sites on a side that an evaluation
of the expectation value of a quantity {\it O}
\begin{equation}
\langle{\it O}\rangle = \frac{\sum_{\{\sigma_{x;\nu}\}}{\it O}
\exp(-\kappa_{\nu\mu}{\cal S})}{\sum_{\{\sigma_{x;\nu}\}}
\exp(-\kappa_{\nu\mu}{\cal S})}
\label{2}
\end{equation}
cannot be performed directly. 
The Monte Carlo method replaces direct evaluation by generation of a sequence
of configurations which simulates an ensemble of states in thermal equilibrium
at inverse temperature $\kappa_{\nu\mu}$.

We have used a heat bath algorithm which creates a Markovian process as
follows: a new value $\sigma_{x;\nu}^{\prime}$ for the link variable
is selected independently on the previous value of $\sigma_{x;\nu}$ in a
stochastic manner with the probability
\begin{equation}
{\cal W}(q_{x;\nu}\rightarrow q_{x;\nu}^{\prime})  
\sim \exp(-\kappa_{\nu\mu}{\cal S}_0(q_{x;\nu}^{\prime})),
\label{3}
\end{equation}
\begin{equation}
{\cal S}_0(q_{x;\nu}^{\prime}) 
= - {\rm Re}\left(\sigma_{x;\nu}^{\prime}\sum_{|\mu|\neq|\nu|}
\kappa_{\nu\mu}\widetilde{\sigma}_{x;\nu\mu}\right),
\label{4}
\end{equation}
\begin{eqnarray}
\widetilde{\sigma}_{x;\nu\mu} &\equiv & \sigma_{x+\nu;\mu}\sigma_{x+\mu;\nu}^*
\sigma_{x;\mu}^*
\label{5} \\
&=& \exp \left(2\pi i (q_{x+\nu;\mu}-q_{x+\mu;\nu}-
q_{x;\mu})/N\right).
\nonumber
\end{eqnarray}
The transition matrix ${\cal W}(q_{x;\nu}\rightarrow q_{x;\nu}^{\prime})$ 
obeys an equation of detailed balance
\begin{equation}
\frac{{\cal W}(q_{x;\nu}\rightarrow q_{x;\nu}^{\prime})}
{{\cal W}(q_{x;\nu}^{\prime}\rightarrow q_{x;\nu})} =
e^{\{-\kappa_{\nu\mu}({\cal S}_0(q_{x;\nu}^{\prime})-
{\cal S}_0(q_{x;\nu}))\}},
\label{6}
\end{equation}
which is a sufficient condition for final distribution to be the Boltzmann
one. A complete cycle through all the link variables of the lattice when the
sites are chosen in a some way (in our case, in a random way) is called as
a Monte Carlo sweep. The detailed balance (\ref{6}) leads to that
the probability to find a configuration $\{\sigma\}$ after $n\rightarrow\infty$
sweeps 
\begin{eqnarray}
{\cal P}(\sigma_{x;\nu}^{\prime}=\sigma) &\sim & 
e^{ -\sum_{|\mu|\neq|\nu|}
\kappa_{\nu\mu}{\rm Re}\exp\left(\sigma+\widetilde{\sigma}_{x;\nu\mu}
\right)},
\nonumber  \\
&& \sum_{q=0,\cdots,N-1} {\cal P} (\sigma_{x;\nu}^{\prime}=\sigma) = 1.
\nonumber
\end{eqnarray}
For the Z(2) gauge group
\begin{eqnarray}
{\cal P}(\sigma_{x;\nu}^{\prime}=1) &=& c\cdot\exp\left(-\sum_{|\mu|\neq|\nu|}
\kappa_{\nu\mu}\widetilde{\sigma}_{x;\nu\mu}\right);
\nonumber \\
{\cal P}(\sigma_{x;\nu}^{\prime}=-1) &=& c\cdot\exp\left(\sum_{|\mu|\neq|\nu|}
\kappa_{\nu\mu}\widetilde{\sigma}_{x;\nu\mu}\right).
\nonumber
\end{eqnarray}
Since the Z(2) gauge group is a discrete group, $-6\leq\sum_{|\mu|\neq|\nu|}
\widetilde{\sigma}_{x;\nu\mu}\leq 6$, the sums
$\sum_\sigma\widetilde{\sigma}_{x;nm}$ and
$\sum_\tau\widetilde{\sigma}_{x;n\tau}$ take the finite number of values,
thereby can play the role of an index in some matrix
\begin{equation}
{\cal T}_{\Sigma_\sigma,\Sigma_\tau} \equiv \exp\left(-\kappa_\sigma
\sum_\sigma \widetilde{\sigma}_{x;nm} -
\kappa_\tau\sum_\tau \widetilde{\sigma}_{x;n\tau}\right)
\label{7}
\end{equation}
\begin{eqnarray}
{\cal P}(\sigma_{x;\nu}^\prime =1)  &\sim &
{\cal T}_{\Sigma_\sigma,\Sigma_\tau},
\label{8} \\
{\cal P}(\sigma_{x;\nu}^\prime =-1) &\sim &
{\cal T}_{-\Sigma_\sigma,-\Sigma_\tau}
\label{9}.
\end{eqnarray}
Neither exponents nor products (expensive computing operations) are
calculated in the course of sweep, they are computed beforehand. Actually
we calculate only the sums $\sum_\sigma$ and $\sum_\tau$  and pick out
the corresponding preliminary calculated exponent in the table (\ref{7})
as the probability to find a configuration $\{\sigma\}$. This algorithm is
applied to every link of the lattice,  giving us
a next in turn configuration from the Boltzmann distribution set.
This procedure is repeated many times, and an estimation of
$\langle{\it O}\rangle$ is obtained.

There are, however, many problems -- both standard and specific for the
discrete groups -- in applying the above algorithm. Due to a discretity
of the Z(2) gauge group, metastable states appear, and 
the system will then remain there for many sweeps if the tunneling 
probability between states is small, especially at large 
$\kappa_{\sigma,\tau}$. This problem can be
hardly solved by increasing the number of sweeps, and in some area of
parameters a new algorithm is required. We have also written
the procedure where new and previous states differ in the value of all
link variables which adjoin the selected link $\sigma_{x;\nu}$.
The correlation between successive configurations then is supposed 
to be weaker. 
Here, we present the results obtained by using a standard
heat bath algorithm. Any procedure gives $\langle{\it O}\rangle$ with
statistical errors. 
The measurements cannot be considered as independent, of course,
because they are obtained on the basis of a sequence of highly correlated
configurations. By taking into account a strong correlation between the
measurements (which is an inevitable consequence of a ``microscopic''
nature of the algorithm -- one link changes at one time), the statistical
error is computed as
\begin{eqnarray}
\varepsilon &=& \sqrt{\chi^2(1+r)/(1-r)},
\label{10} \\
\chi^2      &=&  \sum_i\left({\it O}_i -\bar{\it O}\right)^2/(n-1)n,
\nonumber \\
r           &=&  \frac{\sum_i\left({\it O}_i -\bar{\it O}\right)
                 \left({\it O}_{i-1} -\bar{\it O}\right)}
                 {\sum_i\left({\it O}_i -\bar{\it O}\right)^2},
\nonumber
\end{eqnarray}
where n is the number of samples in the average.
For discrete groups, especially at large $\kappa_{\nu\mu}$, the problem of
correlations becomes to be very sharp, because we can stick at one
configuration, the correlation becomes then practically infinite and
the measured average is questionable.

\section{Results}

We have performed the simulations on the lattice of $N_\sigma^3\times 
N_\tau ~(12^3\times 4)$ size, the total number of sweeps is equal to 
$500 000\times 2$, 
we start from a completely disordered configuration, 
two thousands sweeps were used for thermalization. 
Various quantities such as the space--like (time--like) average plaquette
$\langle P_\sigma\rangle (\langle P_\tau\rangle)$, Wilson loops
$W(I,J)_{\sigma,\tau}$, the Creutz ratio $\chi(I,J) [I=J=2,3,4]$
have been estimated as functions of 
$\beta\equiv\sqrt{\kappa_\sigma\kappa_\tau}=4/g^2$ and $\widetilde{\xi}$.

The temporal $W(2,2)_\tau$ (crosses in Figs) and spatial $W(2,2)_\sigma$ 
(circles in Figs) Wilson loops at 
different $\xi$ are plotted here. When increasing $\beta$, the Wilson loop 
becomes nonzero first in the temporal direction and next in the spatial 
directions -- which is not surprising because $\xi$ enhances the time-like 
plaquettes and suppresses the space-like plaquettes. As is well-known, 
at high temperature (this corresponds to the high anisotropy $\xi\gg 1$) 
the time-like Wilson lines acquire nonzero expectation values, while 
the space-like Wilson lines do not.

Data for the Wilson loops $\langle W_{\sigma,\tau}\left(I\times
J\right)\rangle \simeq \exp\{-\alpha_{\sigma,\tau}\left(\beta,
\widetilde{\xi}\right) \left(I\times J\right) \}$ can be roughly
fitted by
\begin{eqnarray}
\alpha_{\sigma,\tau}\left(\beta,\widetilde{\xi}\right) \simeq 
2\left(1+\varepsilon_{\sigma,\tau}\right) \cdot \left(
\beta_{\sigma,\tau}^{c}-\beta\right) \theta\left(\beta_{\sigma,\tau}^{c}-
\beta\right) \}
\nonumber
\end{eqnarray}
with $0<\varepsilon_{\sigma,\tau}<1;~~\beta_{\sigma,\tau}^{c}=
6\widetilde{\xi}^{v_{\sigma,\tau}};~~v_{\sigma}\sim 0.5,~
v_{\tau}\sim -0.5$ -which does not agree with the condition
obtained for $W_{\sigma,\tau}$ in \cite{burgers} for SU(N)
gauge group. In particular, $W_{\sigma,\tau}$ cannot be fitted 
by the universal function $f\left(\beta,\widetilde{\xi}\right) =
\alpha_{\sigma}\left(\beta,\widetilde{\xi}\right) = 
\xi\left(\beta,\widetilde{\xi}\right) \cdot \alpha_{\tau}
\left(\beta,\widetilde{\xi}\right)$.

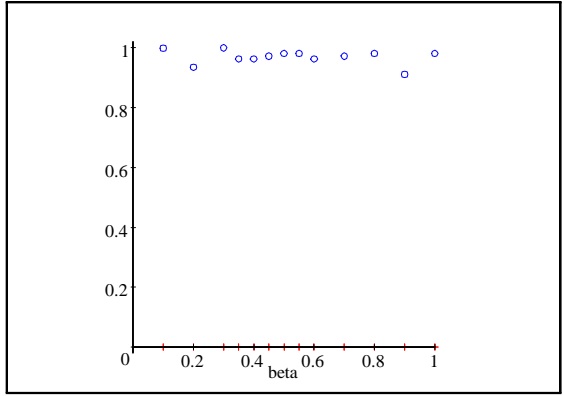
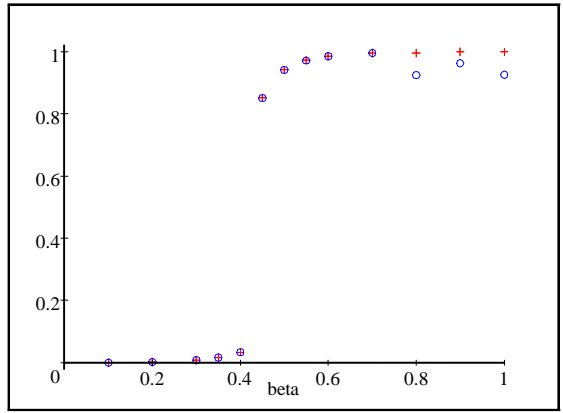
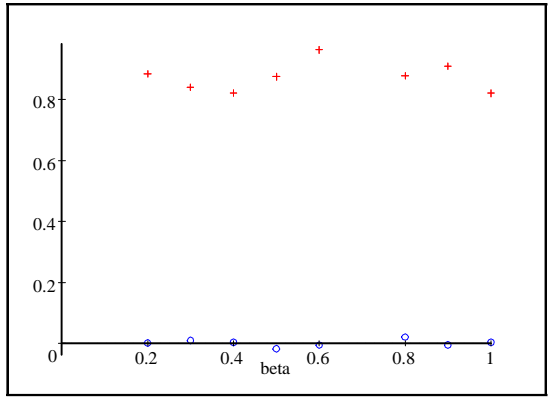
\end{document}